\documentclass[12pt]{article}
\usepackage{amsmath,amssymb} 
\usepackage{subcaption}   

\usepackage{amsfonts}
\usepackage{enumerate}
\usepackage{ytableau}
\usepackage{hyperref}
\usepackage{cite}
\usepackage{bm}
\usepackage{tikz}
\usepackage{blkarray}  
\begin{document}

\def\im{\text{i}}
\def\be{\begin{equation}}
\def\ee{\end{equation}}
\def\bea{\begin{eqnarray}}
\def\eea{\end{eqnarray}}
\def\ba{\begin{array}}
\def\ea{\end{array}}
\def\bd{\begin{displaymath}}
\def\ed{\end{displaymath}}
\def\eg{{\it e.g.~}}
\def\ie{{\it i.e.~}}
\def\Tr{{\rm Tr}}
\def\tr{{\rm tr}}
\def\>{\rangle}
\def\<{\langle}
\def\a{\alpha}
\def\b{\beta}
\def\c{\chi}
\def\del{\delta}
\def\e{\epsilon}
\def\f{\phi}
\def\vf{\varphi}
\def\tvf{\tilde{\varphi}}
\def\g{\gamma}
\def\h{\eta}
\def\j{\psi}
\def\k{\kappa}
\def\l{\lambda}
\def\m{\mu}
\def\n{\nu}
\def\w{\omega}
\def\p{\pi}
\def\q{\theta}
\def\r{\rho}
\def\s{\sigma}
\def\t{\tau}
\def\u{\upsilon}
\def\x{\xi}
\def\z{\zeta}
\def\D{\Delta}
\def\F{\Phi}
\def\G{\Gamma}
\def\J{\Psi}
\def\L{\Lambda}
\def\W{\Omega}
\def\P{\Pi}
\def\Q{\Theta}
\def\S{\Sigma}
\def\U{\Upsilon}
\def\X{\Xi}
\def\nab{\nabla}
\def\pa{\partial}
\newcommand{\lra}{\leftrightarrow}

\newcommand{\bc}{{\mathbb{C}}}
\newcommand{\br}{{\mathbb{R}}}
\newcommand{\bz}{{\mathbb{Z}}}
\newcommand{\bp}{{\mathbb{P}}}

\def\({\left(}
\def\){\right)}
\def\nn{\nonumber \\}

\newcommand{\red}{\textcolor[RGB]{255,0,0}}
\newcommand{\blue}{\textcolor[RGB]{0,0,255}}
\newcommand{\green}{\textcolor[RGB]{0,255,0}}
\newcommand{\cyan}{\textcolor[RGB]{0,255,255}}
\newcommand{\magenta}{\textcolor[RGB]{255,0,255}}
\newcommand{\yellow}{\textcolor[RGB]{255,255,0}}
\newcommand{\sky}{\textcolor[RGB]{135, 206, 235}}
\newcommand{\orange}{\textcolor[RGB]{255, 127, 0}}

\def\ttbar{T$\overline{\text{T}}$ }
\def\Renyi{R$\acute{\text{e}}$nyi }
\def\Poincare{Poincar$\acute{\text{e}}$ }
\def\Banados{Ba$\tilde{\text{n}}$ados }

\title{\textbf{On-shell action of T$\bar{\text{T}}$-deformed Holographic CFTs}}
\vspace{14mm}
\author{Jia Tian\footnote{wukongjiaozi@ucas.ac.cn}}
\date{}
\maketitle

\begin{center}
	{\it
		Kavli Institute for Theoretical Sciences (KITS),\\
		University of Chinese Academy of Science, 100190 Beijing, P.~R.~China
	}
\vspace{10mm}
\end{center}

\makeatletter
\def\blfootnote{\xdef\@thefnmark{}\@footnotetext}  
\makeatother

\begin{abstract}
In this work, we study the holographic dual of the \ttbar deformation following the mixed boundary condition proposal. We point out that a boundary term should be included in the gravity action in the holographic dictionary. In particular, we consider the deformed CFT defined on a sphere (dS) or AdS background and explain the difference between the holographic results and field theory results.
\end{abstract}

\baselineskip 18pt
\newpage

\tableofcontents

\section{Introduction}
The \ttbar deformations \cite{Smirnov:2016lqw,Cavaglia:2016oda} have been studied extensively in recent years due to their remarkable properties \cite{Monica,Jiang:2019epa}. From the perspective of quantum gravity, the two most appealing properties are that the deformed theory is conjectured to be non-local but UV complete and that the deformations have interesting applications to AdS/CFT holography \cite{Maldacena:1997re,Witten:1998qj,Gubser:1998bc}.

The holographic dual of the \ttbar deformation of a holographic  CFT is a bulk gravity theory with mixed boundary conditions for the bulk fields\cite{Guica:2019nzm}. The holographic dictionary for the \ttbar-deformed theories can be summarized as \cite{Monica} \footnote{the notation will become clear in a moment.}
\bea\label{dictionary}
Z_{\text{\ttbar,CFT}}[\gamma^{[\mu]}_{\alpha\beta}]=Z_{\text{grav}}\left[g_{\alpha\beta}^{(0)}+\frac{\mu}{16\pi G_N}g_{\alpha\beta}^{(2)}+\frac{\mu^2}{(16\pi G_N)^2}g_{\alpha\beta}^{(4)}=\gamma^{[\mu]}_{\alpha\beta}\right],
\eea 
where $\mu$ is the deformation parameter. Another well-used holographic proposal is the geometric bulk cut-off proposal \cite{McGough:2016lol,Kraus:2018xrn} (or glue-on proposal \cite{Apolo:2023vnm}) which has some limitations as explained in \cite{Guica:2019nzm}. However, the cut-off proposal is simpler to apply. For example, it is proposed that the entanglement entropy can be computed by the Ryu-Takayanagi (RT) formula \cite{Ryu:2006bv} in the cut-off geometry, and for vacuum states the result agrees with the field theory calculations \cite{Donnelly:2018bef}. But a disagreement has also been found in \cite{Chen:2018eqk} when they considered the deformed entanglement entropy of a thermal state.  In this paper, we want to argue that the RT formula does not give the full answer of the entanglement entropy because there is an extra term in the gravity action.

The key observation of us is that in the dictionary \eqref{dictionary} we should add a proper boundary term to the gravity action\footnote{Perhaps this is already hidden written in\cite{Guica:2019nzm}.}. The presence of this boundary term is natural considering that the double trace deformation not only shifts the source but also shifts the generating function \cite{Klebanov:1999tb,Witten:2001ua}. We will compute several examples explicitly to confirm this observation. Shifting the action by a universal constant will only change the entanglement entropy by some constant which can be absorbed into the UV cut-off. However, if the entanglement entropy does not depend on the UV cut-off or the constant is not universal but state-dependent then this constant should have physical meaning and the correct RT formula should capture it.

Usually \ttbar deformation is only well defined for the theories on flat spacetime. For example, the important factorization property of the \ttbar operator will lose on a general curved background \cite{Jiang:2019tcq}. However for holographic CFTs, in the large $N$ limit, composite operators always factorize and the dictionary \eqref{dictionary} applies to arbitrary 2-dimensional (2d) backgrounds even though constructing the general bulk dual is very challenging. By restricting our considerations to the vacuum state and a curved background with a constant Ricci curvature, we find that the construction of the bulk dual can be simplified to solve the well-known and integrable Liouville equation. The partition function of the \ttbar-deformed CFT on a sphere in the vacuum state has been derived in \cite{Donnelly:2018bef} by solving the flow equation directly. Our holographic result differs from it by a universal Weyl anomaly constant which is proportional to the Euler character of the background and a $\mu$--dependent constant. From a deformation perspective, our holographic result is favored because it has an undeformed limit. In contrast, the partition function derived in \cite{Donnelly:2018bef} diverges in the limit $\mu\rightarrow 0$.
The Weyl anomaly constant is also important. It turns out that it will cancel out the UV divergence in the entanglement entropy. From our point of view, this cancellation explains why in \cite{Donnelly:2018bef} the entanglement entropy is UV finite.  We further consider the AdS background. Because the AdS space is not compact the on-shell action is divergent however the holographic entanglement entropy can be computed in the same way and the result is similar to the one in the sphere background case.

\section{Holographic dictionary of \ttbar deformed theories}
\renewcommand{\theequation}{2.\arabic{equation}}
\setcounter{equation}{0}
In this section, we briefly review the derivation of the holographic dictionary \eqref{dictionary}. The derivation \cite{Guica:2019nzm} is based on the variational principle and the holographic dual of the double-trace deformation \cite{Klebanov:1999tb,Witten:2001ua}.  
Our convention and notation are summarized in Appendix \ref{Convention}.

\subsection{Setup}
The \ttbar deformation of 2d field theory with action $S_{\text{CFT}}$ is defined by the flow equation
\bea
\frac{dS^{[\mu]}_{\text{CFT}}}{d\mu}=\int d^2 x \sqrt{\gamma}\, \text{T}\bar{\text{T}}^{[\mu]},
\eea 
where in principle the 2d background can be arbitrary. In the linear order, the deformation is just a double trace deformation
\bea \label{fieldaction}
S_{CFT}^{[\mu]}=S_{CFT}+\mu \int d^2 x \sqrt{\gamma}\, \text{\ttbar}+\mathcal{O}(\mu^2).
\eea 
The double trace deformation does two things to the generating function. It shifts the source by the expectation value of the dual operator and it shifts the generating function (or the on-shell action) by \textit{subtracting} the double trace operator, \ie
\bea \label{doubletr}
W^{[\mu]}=W_0-\mu \int d^2 x \sqrt{\gamma}\, \text{\ttbar}.
\eea 
Using the defining property of the generating function $\delta W=\frac{1}{2}\int d^2 x\sqrt{\gamma}T_{\alpha\beta}\delta\gamma^{\alpha\beta}$, we can obtain a flow equation 
\bea 
\frac{1}{2}\partial_\mu\(\int d^2x \sqrt{\gamma^{[\mu]}}T_{\alpha\beta}^{[\mu]}\delta \gamma_{\alpha\beta}^{[\mu]}\)=-\delta\(\int d^2x \sqrt{\gamma^{[\mu]}}\text{T}\bar{\text{T}}^{[\mu]}\),
\eea 
which can be solved by \cite{Guica:2019nzm}
\bea 
&&\gamma_{\alpha\beta}^{[\mu]}=\gamma_{\alpha\beta}^{[0]}+\frac{1}{2}\mu \hat{T}_{\alpha\beta}^{[0]}+\frac{1}{16}\mu^2\hat{T}_{\alpha\rho}^{[0]}\hat{T}_{\sigma\beta}^{[0]}{\gamma^{[0]}}^{\rho\sigma},\label{metricflow}\\
&&\hat{T}^{[\mu]}_{\alpha\beta}=\hat{T}^{[0]}_{\alpha\beta}+\frac{1}{4}\mu\hat{T}_{\alpha\rho}^{[0]}\hat{T}_{\sigma\beta}^{[0]}{\gamma^{[0]}}^{\rho\sigma},\label{energyflow}
\eea 
where $\hat{T}_{\alpha\beta}=T_{\alpha\beta}-\gamma_{\alpha\beta}T$. The proposal of \cite{Guica:2019nzm} is that $\gamma^{[\mu]}_{\alpha\beta}$ and $T^{[\mu]}_{\alpha\beta}$ are the sources and dual operators of the deformed holographic theory which is still an asymptotic AdS$_3$ gravity theory. The general asymptotic AdS$_3$ solution can be written in the Fefferman-Graham gauge like \cite{Skenderis:1999nb}
\bea 
ds^2=g_{\alpha\beta}(\rho,x^\alpha)dx^\alpha dx^\beta+\frac{d\rho^2}{4\rho^2},\quad g_{\alpha\beta}(\rho,x^\alpha)=\frac{g_{\alpha\beta}^{(0)}}{\rho}+g_{\alpha\beta}^{(2)}+\rho g_{\alpha\beta}^{(4)},
\eea 
where $g^{(2)}$ corresponds to the initial expectation value of the CFT operator
\bea \label{energymap}
\hat{T}^{[0]}_{\alpha\beta}=\frac{1}{8\pi G_N}g_{\alpha\beta}^{(2)}.
\eea 
The solution \eqref{metricflow} implies that the deformed boundary metric is
\bea 
\gamma_{\alpha\beta}^{[\mu]}=g_{\alpha\beta}^{(0)}+\frac{\mu}{16\pi G_N}g_{\alpha\beta}^{(2)}+\frac{\mu^2}{(16\pi G_N)^2}g_{\alpha\beta}^{(4)}=\rho_c g_{\alpha\beta}(\rho_c),\quad \rho_c=\frac{\mu}{16\pi G_N },
\eea 
which gives the holographic dictionary \eqref{dictionary} as proposed in \cite{Guica:2019nzm}. The dictionary seems to suggest that the on-shell action of the holographic theory is simply
\bea 
I_{\text{Euclidean}}^{[\mu]}=I_{\text{Euclidean}}^{[0]}\(\gamma^{[\mu]}_{\alpha\beta}=g_{\alpha\beta}^{(0)}+\frac{\mu}{16\pi G_N}g_{\alpha\beta}^{(2)}+\frac{\mu^2}{(16\pi G_N)^2}g_{\alpha\beta}^{(4)}\).
\eea 
But we will show that it is not correct. The proper on-shell action should be 
\bea 
I_{\text{on-shell}}^{[\mu]}&=&I_{\text{Euclidean}}^{[0]}\(\gamma^{[\mu]}_{\alpha\beta}=g_{\alpha\beta}^{(0)}+\frac{\mu}{16\pi G_N}g_{\alpha\beta}^{(2)}+\frac{\mu^2}{(16\pi G_N)^2}g_{\alpha\beta}^{(4)}\)-\mu\int\sqrt{\gamma^{[\mu]}}\text{T}\bar{\text{T}}^{[\mu]} \nonumber\\
&\equiv &I_{\text{bulk}}^{[\mu]}+I_{\text{bdy}}^{[\mu]},\label{onshell}
\eea 
which is similar to the one \eqref{doubletr} of double trace deformation. The second term is a surface integral and it will not modify the bulk equation of motion. But it contributes to the on-shell action and it is necessary to include it if we want to compute the correlation functions or entanglement entropy correctly. For example, it will potentially modify the description of the RT surface which we also comment on below. Using \eqref{metricflow} and \eqref{energyflow} one can show that $\sqrt{\gamma}\text{\ttbar}$ is invariant under the flow so the boundary term in \eqref{onshell} can also be written as
\bea 
-\mu\int\sqrt{\gamma^{[\mu]}}\text{T}\bar{\text{T}}^{[\mu]}=-\mu \int \sqrt{g^{[0]}}\text{T}\bar{\text{T}}^{[0]}.
\eea

\subsection{The On-shell action}
Choosing the conformal gauge, the boundary metric can be written as
\bea 
g_{ij}^{(0)}dx^i dx^j=e^\phi dy d\bar{y},
\eea 
and complete 3d metric is given by \cite{Skenderis:1999nb} \footnote{ The AdS radius $l$ is set to be $1$.}
\bea \label{gme}
&&ds^2=\frac{d\rho^2}{4\rho^2}+\frac{1}{\rho}e^\phi dy d\bar{y}+\frac{1}{2}\mathcal{T}_\phi dy^2+\frac{1}{2}\bar{\mathcal{T}}_\phi d\bar{y}^2+\frac{1}{4}R_\phi dy d\bar{y}\nn
&&\quad +\frac{1}{4}\rho e^{-\phi}(\mathcal{T}_\phi dy+\frac{1}{4}R_\phi d\bar{y})(\bar{\mathcal{T}}_\phi d\bar{y}+\frac{1}{4}R_\phi d{y}),
\eea 
where
\bea 
\mathcal{T}_\phi=\partial_y^2\phi-\frac{1}{2}(\partial_y \phi)^2+L(y),\quad \bar{\mathcal{T}}_\phi=\bar{\partial}_y^2\phi-\frac{1}{2}(\bar{\partial}_y \phi)^2+\bar{L}(\bar{y}),\quad R_\phi=4\partial_y \bar{\partial}_y \phi
\eea 
and the boundary Ricci scalar for the metric $g_{ij}^{(0)}$ is $R^{(0)}=-e^{-\phi}R_{\phi}$. $L(y)$ and $\bar{L}(\bar{y})$ are two arbitrary functions that characterize different states of the holographic CFT. This bulk solution is the most general AdS$_3$  solution and it is a generalization of the well-known \Banados geometry which has a flat boundary metric.
The vacuum state corresponds to the case when $L(y)=\bar{L}(\bar{y})=0$ and the vacuum metric can be mapped to the \Poincare metric by \footnote{For non-trivial $L(y)$ and $\bar{L}(\bar{y})$ the metric can also be mapped to \Poincare metric since AdS$_3$ gravity has no local degrees of freedom but the transformation is much more complicated.}
\bea \label{topoincare}
\frac{1}{\eta}=\rho^{-1/2}e^{\phi/2}+\frac{1}{4}\rho^{1/2}e^{-\phi/2}|\partial_y \phi|^2,\quad z=y+\frac{1}{2}\frac{\rho e^{-\phi} \bar{\partial}_y \phi}{1+\frac{1}{4}\rho e^{-\phi}|\partial_y \phi|^2},
\eea 
with the resulting metric
\bea 
ds^2=\frac{d\eta^2+ dz d\bar{z}}{\eta^2}.
\eea 
The 3d Euclidean AdS Einstein  gravity has the action
\bea 
I_E^{[0]}=\frac{1}{16\pi G_N}\(-\int_B \sqrt{h}(R+2)-2\int_{\partial B} \sqrt{\gamma}K+2\int_{\partial B} \sqrt{\gamma}\).
\eea 
Here $\partial B$ is the UV regulator surface which is usually chosen to be at
\bea \label{uv}
 \rho_{UV}=\delta^2,
\eea 
which implies
\bea 
&&\sqrt{h}=\sqrt{g^{(0)}}\(\frac{1}{2\rho^2}+\frac{e^{-\phi}R_\phi}{8\rho}+\frac{e^{-2\phi}}{128}(R_\phi^2-16\mathcal{T}_\phi \bar{\mathcal{T}}_\phi)\),\\
&&\sqrt{\gamma}=\sqrt{g^{(0)}}\(\frac{1}{\delta^2}+\frac{e^{-\phi}R_\phi}{4}+\frac{e^{-2\phi}}{64}(R_\phi^2-16\mathcal{T}_\phi \bar{\mathcal{T}}_\phi)\delta^2\),\\
&&\sqrt{\gamma}K=\sqrt{g^{(0)}}\(\frac{2}{\delta^2}-\frac{e^{-2\phi}}{32}(R_\phi^2-16\mathcal{T}_\phi \bar{\mathcal{T}}_\phi)\delta^2\).
\eea
 The 3d metric degenerates at the positions where $\sqrt{h}=0$ which has two solutions 
\bea 
\rho_\pm=\frac{8 e^\phi}{-R_\phi\pm 4\sqrt{\mathcal{T}_\phi \bar{\mathcal{T}}_\phi}}.
\eea 
When $\rho_+>0$, we should only include the spacetime below the curve $\rho_H=\rho_+$. Then the on-shell action is equal to
\bea \label{onshell1}
I_E^{[0]}=-\frac{1}{16\pi G_N} \int_{\partial B}\sqrt{g^{(0)}}\(\frac{1}{2}R^{(0)}(1+\log{\rho_H})+2\sqrt{\mathcal{T}_\phi \bar{\mathcal{T}}_\phi}\)+\frac{c}{6}\chi(\partial B)\log\delta,
\eea 
where we have used the relation $c=\frac{3}{2G_N}$ and $\chi(\partial B)$ denotes the Euler character of the boundary manifold $\partial B$. The second term is divergent but universal which captures the Weyl anomaly. Sometimes in the literature, the second term is ignored so that the on-shell action is finite. But later we will see that this term will contribute to the entanglement entropy.
When the boundary metric is flat and $\chi=0$, the on-shell action reduces to 
\bea \label{flatonshell}
I_E^{[0]}=-\frac{c}{12\pi}\int \sqrt{g^{(0)}}{\sqrt{\mathcal{T}_\phi \bar{\mathcal{T}}_\phi}}.
\eea  
To compute the first term in the deformed on-shell action \eqref{onshell}, we only need to choose the physical metric $\gamma^{[\mu]} $ on which the deformed theory is defined, solve the bulk metric according to the holographic dictionary and then substitute the results into the general formula \eqref{onshell1}. To compute the boundary term we use the holographic dictionary \eqref{energymap} and the result is
\bea 
-\mu\int \sqrt{g^{(0)}}\text{T}\bar{\text{T}}^{[0]}=\frac{\mu}{4096 G_N^2\pi^2}\int \sqrt{g^{(0)}} e^{-2\phi}\(R_\phi^2-16\mathcal{T}_\phi \bar{\mathcal{T}}_\phi\),
\eea 
which in the flat limit reduces to
\bea 
-\frac{\mu c^2}{576\pi^2}\int \sqrt{g^{(0)}} \mathcal{T}_\phi \bar{\mathcal{T}}_\phi.
\eea 

\section{Examples}
\renewcommand{\theequation}{3.\arabic{equation}}
\setcounter{equation}{0}
\subsection{Torus background: the thermal state in flat spacetime}
The thermal state in flat spacetime corresponds to choosing the following  parameters
\bea 
\phi=0,\quad \mathcal{T}_\phi=\bar{\mathcal{T}}_\phi=2L_0,
\eea 
where $L_0$ is some constant. The bulk geometry is the BTZ black hole whose metric is
\bea \label{btz}
ds^2=\frac{d\rho^2}{4\rho^2}+\frac{(1+L_0 \rho)^2 dx^2+(-1+L_0\rho)^2 d\tau^2}{\rho}
\eea 
and the asymptotic boundary manifold is a torus with the spatial and thermal period being $w_0,\beta_0=\pi/\sqrt{L_0}$, respectively. Using \eqref{flatonshell}, we can get the on-shell action
\bea \label{btzonshell}
I_E^{[0]}=-\frac{c}{12\pi} \omega_0\beta_0 (2L_0)=-\frac{c\pi}{6}\frac{\omega_0}{\beta_0}.
\eea 
When the deformed metric is flat so is the boundary metric and these two metrics are related by a coordinate transformation \cite{Guica:2019nzm} which is reminiscent of the dynamical coordinate transformation interpretation of \ttbar deformation introduced in \cite{Dubovsky:2012wk,Dubovsky:2017cnj}.

According to the dictionary \eqref{dictionary}, this bulk solution \eqref{btz} is dual to the deformed theory defined on the 2d torus with metric\footnote{Note that here $\rho_c$ can be both positive and negative.}
\bea \label{deformBTZ}
ds^2=(L_0\rho_c+1)^2dx^2+(L_0\rho_c-1)^2d\tau^2\equiv dX^2+dY^2,
\eea 
whose spatial and thermal period $w,\beta$ are related by $w_0,\beta_0$ via
\bea \label{relation}
\beta_0=\frac{1}{2} \left(\beta +\sqrt{\beta ^2+4 \pi ^2 {\rho_c}}\right),\quad \omega_0=\frac{1}{2} \omega \left(\frac{\beta }{\sqrt{\beta ^2+4 \pi ^2 \rho_c}}+1\right).
\eea 
Therefore the bulk part of the deformed on-shell action is simply given by substituting \eqref{relation} into \eqref{btzonshell}
\bea 
I_{\text{bulk}}^{[\mu]} &=&-\frac{c\pi}{6}\frac{\omega_0}{\beta_0}=-\frac{c\pi}{6}\frac{\omega}{\sqrt{\beta^2+4\pi^2\rho_c}} \\
&=&-\frac{c\pi}{6}\frac{\omega}{\beta}+\frac{c\pi^3\omega\rho_c}{3\beta^3}+\mathcal{O}(\rho_c^2),
\eea 
and the boundary part is 
\bea 
I_{\text{bdy}}^{[\mu]}&=&-\mu\int \sqrt{\gamma^{[0]}} \langle \text{\ttbar}\rangle=-\mu \int \(\frac{L_0}{8\pi G_N}\)^2=-\frac{\mu \omega_0\beta_0}{(8\pi G)^2}\(\frac{\pi^2}{\beta_0^2}\)^2 \nonumber\\
&=&-\frac{2c\rho_c  \pi^3 w}{3\sqrt{4\pi^2\rho_c+\beta^2}(\beta+\sqrt{4\pi^2\rho_c+\beta^2})^2}=-\frac{c\pi^3w\rho_c}{6\beta^3}+\mathcal{O}(\rho_c^2).
\eea 
Adding these two terms together we get the final deformed on-shell action
\bea 
I_{\text{on-shell}}^{[\mu]}&=&\frac{c\omega(\beta-\sqrt{\beta^2+4\pi^2\rho_c})}{12\pi \rho_c}=-\frac{c\pi \omega}{6\beta}+\frac{c\pi^3\omega \rho_c}{6\beta^3}+\mathcal{O}(\rho^2_c)\\
&=&I_{\text{on-shell}}^{[0]}+\mu\int \sqrt{\gamma^{[0]}} \langle \text{\ttbar}\rangle+\mathcal{O}(\rho^2_c),
\eea 
which matches the perturbative results of the field theory.

To verify the proposal of the on-shell action. Let us consider the entanglement entropy of the whole spatial circle. Using the replica trick, the \Renyi entropy can be computed as
\bea 
S_n&=&\frac{1}{1-n}\log \frac{Z_n}{Z_1^n}=\frac{1}{1-n}\log \frac{e^{-I^{[\mu]}_{\text{on-shell}}(n\beta)}}{e^{-n I^{[\mu]}_{\text{on-shell}}(\beta)}}=\frac{c\omega\(n\sqrt{4\pi^2\rho_c+\beta^2}-\sqrt{4\pi^2\rho_c+n^2\beta^2}\)}{12(n-1)\pi\rho_c}\nonumber\\
&=&\frac{c\pi \omega}{3\sqrt{4\pi^2\rho_c+\beta^2}}-\frac{(n-1)c\pi\omega\beta^2}{6 (4\pi^2\rho_c+\beta^2)^{3/2}}+\mathcal{O}((n-1)^2).
\eea 
Therefore the entanglement entropy is 
\bea
S_1= \frac{c\pi \omega}{3\sqrt{4\pi^2\rho_c+\beta^2}}=\frac{c\pi w_0}{3\beta_0}=\frac{\gamma}{4G_N},
\eea 
which agrees with the RT formula.

\subsection{Conical background: the primary state in flat spacetime}
Let us start from the vacuum \Poincare AdS$_3$ metric 
\bea 
ds^2=\frac{dw d\bar{w }+dz^2}{z^2},
\eea 
and consider the conformal transformation $w=y^n,\bar{w}=\bar{y}^n$. Using the \Banados map \cite{Banados:1998gg} we obtain a excited bulk geometry in the form \eqref{gme} with
\bea 
\phi=0,\quad \mathcal{T}_\phi=\frac{n^2-1}{2y^2}\equiv\frac{2a^2}{y^2},\quad \bar{\mathcal{T}}_\phi=\frac{2a^2}{\bar{y}^2}.
\eea 
In the radial coordinates the 3d metric is 
\bea \label{conical}
ds^2=\frac{d\rho^2}{4\rho^2}+\frac{1}{\rho}\((1+\frac{a^2 \rho}{r^2})^2 dr^2+(r-\frac{a^2 \rho}{r})^2 d\theta^2\),
\eea 
which degenerates at $\rho_H=r^2/a^2$, so we only need to consider the spacetime below it. The  boundary manifold has a conical singularity at $r=0$ since $\theta\sim \theta+2\pi/n$. Using \eqref{flatonshell} we can get the on-shell action 
\bea 
I_E^{[0]}=-\frac{c}{12\pi}\int_{\epsilon_0}^{\Lambda_0} rdr\int_0^{\frac{2\pi}{n}}d\theta\, \frac{2a^2}{r^2}=-\frac{ca^2}{3n}\log\frac{\Lambda_0}{\epsilon_0}.
\eea 
The bulk solution \eqref{conical} is dual to the deformed theory which is defined on the 2d disk with metric
\bea 
ds^2=(1+\frac{a^2\rho_c}{r^2})^2dr^2+(r-\frac{a^2\rho_c}{r})^2d\theta^2\equiv dR^2+R^2d\theta^2,
\eea 
where
\bea 
r=\frac{1}{2}\(\sqrt{R^2+4a^2\rho_c}+R\).
\eea 
Then the deformed on-shell action is given by 
\bea 
&&I_{\text{on-shell}}^{[\mu]}=-\frac{ca^2}{3n}\log\frac{\Lambda_0}{\epsilon_0}-\mu\int \sqrt{\gamma^{[0]}}\langle \text{\ttbar}\rangle\nonumber \\
&&=-\frac{ca^2}{3n}\log\frac{\Lambda+\sqrt{4a^2\rho_c+\Lambda^2}}{\epsilon+\sqrt{4a^2\rho_c+\epsilon^2}}+\frac{2a^4c \rho_c   }{3n}\left(\frac{1}{\left(\Lambda +\sqrt{4 a^2 \rho_c+\Lambda ^2}\right)^2}-\frac{1}{\left(\epsilon +\sqrt{4 a^2 \rho_c+\epsilon ^2}\right)^2}\right)\nonumber \\
&&=\frac{a^2 c}{3n}\log\frac{\epsilon}{\Lambda}+\frac{a^4c\rho_c}{6n}\(\frac{1}{\epsilon^2}-\frac{1}{\Lambda^2}\)+\mathcal{O}(\rho_c^2)=I_{\text{on-shell}}^{[0]}+\mu\int \sqrt{\gamma^{[0]}}\langle \text{\ttbar}\rangle+\mathcal{O}(\rho_c^2)
\eea 
as desired. 
\subsection{Non-flat boundary}
The general bulk solution \eqref{gme} depends on three arbitrary functions $\phi(y,\bar{y}),L(y)$ and $\bar{L}(\bar{y})$. It is either dual to a CFT defined on the boundary with metric $g_{\mu\nu}^{[0]}$ or dual to a \ttbar-deformed QFT defined on a 2d manifold with metric 
\bea \label{dgem}
\gamma_{\alpha\beta}^{[\mu]}dx^\alpha dx^\beta &=&\frac{(8-R^{(0)} \rho_c)^2}{64}e^{\phi}\(dy+\frac{4\rho_c e^{-\phi}\bar{\mathcal{T}}_\phi}{8-R^{(0)} \rho_c}d\bar{y}\)\(d\bar{y}+\frac{4\rho_c e^{-\phi}\mathcal{T}_\phi}{8-R^{(0)} \rho_c}d{y}\) \\
&\equiv& e^{\sigma}dU d\bar{U},
\eea 
where the two metrics are related by a coordinate transformation and a Weyl transformation \cite{Caputa:2020lpa} as
\bea 
&&\lambda(y,\bar{y})\(dy+\frac{4\rho_c e^{-\phi}\bar{\mathcal{T}}_\phi}{8-R^{(0)} \rho_c}d\bar{y}\)=dU,\quad \bar{\lambda}(y,\bar{y})\(d\bar{y}+\frac{4\rho_c e^{-\phi}\mathcal{T}_\phi}{8-R^{(0)} \rho_c}d{y}\)=d\bar{U},\\
&&e^{\sigma}=\frac{(8-R^{(0)} \rho_c)^2}{64 \lambda\bar{\lambda}}e^{\phi}.
\eea 
The two functions $\lambda$ and $\bar{\lambda}$ are integrating factors that are usually hard to determine but they do exist according to the theory of differential equations. In particular, they are not unique because the conformal gauge is not unique. If $\lambda,\bar{\lambda}$ are proper integrating factors then $F(U(y,\bar{y}))\lambda$ and $\bar{F}(\bar{U}(y,\bar{y}))\bar{\lambda}$ are also good integrating factors. When the boundary manifold is flat \ie $\phi=0$  we can choose $\lambda=\bar{\lambda}=1$ such that
\bea 
dU=dy+\frac{1}{2}\bar{\mathcal{T}}_\phi(\bar{y})d\bar{y},\quad d\bar{U}=d\bar{y}+\frac{1}{2}{\mathcal{T}}_\phi({y})d{y},\quad e^\sigma=\frac{1}{8},
\eea 
which implies that metric $\gamma_{\alpha\beta}$ is also flat. 
In general, if we choose an arbitrary  $\gamma_{\alpha\beta}$  it is very challenging to construct the bulk solution. 
Below we will consider some special cases where the metric $\gamma_{\alpha\beta}$ is maximally symmetric. 

\subsection{Sphere background in the vacuum state}
To fix $\gamma_{\alpha\beta}^{[\mu]}$ to be a sphere, one should start from the general metric \eqref{dgem} and solve the Einstein equation $R^{[\mu]}=-\Lambda,\, \Lambda<0$. However, for the special case $L=\bar{L}=0$ which corresponds to the ground state, it turns out that if the boundary metric $g_{\alpha\beta}^{[0]}$ is a sphere then $\gamma_{\alpha\beta}^{[\mu]}$ is also a sphere but with a different radius. The boundary Einstein equation $R^{(0)}=-\Lambda_0$ is much simpler to solve because it is just the famous Liouville equation
\bea 
4\partial_y \partial_{\bar{y}}\phi-e^\phi \Lambda_0=0,
\eea 
whose solution is 
\bea 
\phi=-2\log(1+y\bar{y})+\log(\frac{-8}{\Lambda_0}),\quad \Lambda_0<0.
\eea 
The resulting 3d metric can be written as
\bea \label{3dva}
ds^2=\frac{d\rho^2}{4\rho^2}+\frac{(8+\Lambda_0 \rho)^2}{(-8\Lambda_0)(1+\kappa^2)^2\rho}(d\kappa^2+\kappa^2d\varphi^2),\quad y=\kappa e^{\im \varphi},
\eea 
which can be transformed into the AdS global coordinates
\bea 
ds^2={d\tilde{\rho}^2}+\sinh^2\tilde{\rho}(d\theta^2+\sin^2\theta d\varphi^2)
\eea 
via the coordinate transformation
\bea 
\theta=2\arctan \kappa,\quad \sinh\tilde{\rho}=\frac{8+\Lambda_0 \rho}{\sqrt{-32 \Lambda_0 \rho}}.
\eea 
Using the dictionary \eqref{dictionary} one can easily find that deformed metric 
\bea 
ds^2=-\frac{(8+\Lambda_0 \rho_c)^2}{8\Lambda_0(1+y\bar{y})^2}dyd\bar{y}\equiv -\frac{8}{\Lambda}\frac{dyd\bar{y}}{(1+y\bar{y})^2},
\eea
indeed describes a sphere with a different radius given by
\bea\label{lambda}
\Lambda_0=-\frac{8(\rho_c \Lambda-4+2\sqrt{4-2\rho_c \Lambda})}{\rho_c^2\Lambda}.
\eea 
Ignoring the universal Weyl anomaly divergent term for a moment, the on-shell action \eqref{onshell1} for this geometry is 
\bea 
&&I_{\text{bulk}}^{[\mu]}=-\frac{c}{6}\(1+\log\frac{8}{-\Lambda_0}\)=\frac{c}{6}\(1+\log\frac{\rho_c^2\Lambda}{\Lambda \rho_c-4+2\sqrt{4-2\Lambda \rho_c}}\),\\
&&I_{\text{bdy}}^{[\mu]}=\frac{\mu}{4096 G_N^2\pi^2}\int \sqrt{g^{[0]}}e^{-2\phi}R_\phi^2=-\frac{c\Lambda_0 \rho_c}{48},\\
&&I_{\text{on-shell}}^{[\mu]}=I_{\text{bulk}}^{[\mu]}+I_{\text{bdy}}^{[\mu]}=-\frac{c}{6}\(1+\log\frac{8}{-\Lambda}\)+\frac{c\Lambda \rho_c}{48}+\mathcal{O}(\rho_c^2).
\eea 
Comparing with the results \eqref{fieldsphere} of the field theory we see that they are different by a $\mu$-dependent constant \footnote{Note that the relation between the Ricci scalar and the radius of the sphere is $\Lambda \to -\frac{2}{r^2}$.}
\bea 
I_{\text{on-shell}}^{[\mu]}-S_{QFT}^\mu=\frac{c}{6}\log \frac{24\pi}{c\mu}.
\eea 
This is because instead choosing the limit condition $S_{QFT}^\mu(r=0)=0$, we have chosen a different scheme 
\bea 
I_{\text{on-shell}}^{[\mu]}(\Lambda=-\infty)=\frac{c}{6}\log\frac{24\pi}{c\mu}.
\eea 
The decision to set $S_{QFT}^\mu(r=0)=0$ in \cite{Donnelly:2018bef} is based on the assumption that the action should become zero when the sphere collapses to a point. From our perspective, it is not a favorable option as the action $S_{QFT}^\mu$ lacks an undeformed limit. In our approach, there is an ambiguity to choose the UV cut-off when we define the regulated on-shell action and we have specially chosen the one which has an undeformed limit. By using the replica trick, one can show that shifting the on-shell action will shift the entanglement entropy as
\bea 
I_{\text{on-shell}}\rightarrow I_{\text{on-shell}}+\alpha,\quad S_A\rightarrow S_A-\alpha.
\eea 
Usually, this constant can be absorbed into the UV cut-off when the constant is finite. When $\alpha$ is also infinite, the situation becomes very subtle. For example, if we also include the Weyl anomaly contribution, the total action should be
\bea 
I_{\text{on-shell,Weyl}}^{[\mu]}=I_{\text{on-shell}}^{[\mu]}+\frac{c}{6}\log \delta^2.
\eea 
Note that if the entanglement entropy of the theory with the action $ I_{\text{on-shell,Weyl}}^{[\mu]} $ has a universal UV divergence 
\bea 
\frac{c}{3}\log\frac{1}{\delta}
\eea 
then the theories with the action $ I_{\text{on-shell}}^{[\mu]} $ and also $S_{QFT}^\mu$ will not have it because it is canceled by the universal Weyl anomaly. This is the reason why in \cite{Donnelly:2018bef}, a UV finite entanglement entropy is obtained. Have figured out how the entanglement entropy for different actions is related to each other, let us compute the deformed entanglement entropy. It turns out that the RT formula gives the correct result up to a constant. 

It is  useful to compute the undeformed entanglement entropy first.
Let us consider a single interval with two endpoints. Using the isometry of the sphere we can always choose them to be 
$(\rho,y,\bar{y})=(\delta^2,0,0)$ and $(\rho,y,\bar{y})=(\delta^2,\tan\frac{\theta_0}{2},\tan\frac{\theta_0}{2})$.
Then using the RT formula we can find the entanglement entropy
\bea 
S_A=\frac{\gamma}{4G_N}=\frac{c}{6}\log\(\frac{8\sin^2\frac{\theta_0}{2}}{-\Lambda_0\delta^2}\).
\eea 
The geodesic length can be easily found by transforming the two points into the \Poincare coordinates via \eqref{topoincare}. Then the deformed entanglement entropy should be
\bea 
S_A^{(\mu)}&=&\frac{c}{6}\log\(\frac{8\sin^2\frac{\theta_0}{2}}{-\Lambda_0\delta^2}\)=\frac{c}{6}  \log \left(\frac{\Lambda  \rho _c^2 \sin ^2\left(\frac{\theta _0}{2}\right)}{\delta ^2 \left(2 \sqrt{4-2 \Lambda  \rho _c}+\Lambda  \rho _c-4\right)}\right),\\
&=&\frac{c}{6}\log\frac{8\sin^2\frac{\theta_0}{2}}{-\Lambda\delta^2}-\frac{c^2\Lambda \mu}{576\pi}+\mathcal{O}(\mu^2).
\eea 
For the special case when $\theta_0=\pi$, the result coincides with one in \cite{Donnelly:2018bef} if we take accounts of the shift due to Weyl anomaly and the shift between $ I_{\text{on-shell}}^{[\mu]} $ and $S_{QFT}^\mu$. The other thing we want to emphasize is that our result is valid for both signs of the deformation parameter $\mu$. 
\subsection{\Poincare disk background in the vacuum state}
The situation is very similar to the one with the sphere background. For negative Ricci curvature, the solution of the Liouville equation is   
\bea 
\phi=-2\log(1-y\bar{y})+\log(\frac{8}{\Lambda_0}),\quad \Lambda_0>0
\eea 
and the deformed metric is 
\bea 
\gamma_{ij}^{(\mu)}dx^i dx^j=\frac{(8+\Lambda_0 \rho_c)^2}{8\Lambda_0(1-y\bar{y})^2}dyd\bar{y}\equiv \frac{8}{\Lambda}\frac{dyd\bar{y}}{(1-y\bar{y})^2},
\eea 
with
\bea 
\Lambda_0=-\frac{8 \left(2 \sqrt{4-2 \Lambda  \rho _c}+\Lambda  \rho _c-4\right)}{\Lambda  \rho _c^2}.
\eea 
Since the $AdS$ space is not compact, the on-shell action has an IR divergence. So let us focus on the holographic entanglement entropy. For the interval with endpoints at $(y,\bar{y})=(0,0)$ and $(a,a)$, the entanglement entropy is 
\bea 
S_A^{(\mu)}&=&\frac{c}{6}\log\frac{8a^2}{(\Lambda_0(1-a^2)\delta^2)}=\frac{c}{6}  \log \left(\frac{a^2 \Lambda  \rho _c^2}{\left(a^2-1\right) \delta ^2 \left(2 \sqrt{4-2 \Lambda  \rho _c}+\Lambda  \rho _c-4\right)}\right).
\eea

\section{Conclusion and Discussion}
In this paper, we want to emphasize that in the holographic dictionary \eqref{dictionary} of the \ttbar-deformed CFTs the gravity action should also include an additional boundary term. This appearance of this boundary term inherits from the fact that the leading order \ttbar deformation is a double trace deformation. We confirm this observation in some explicit examples including the BTZ, conical spacetime and the vacuum state of a holographic CFT defined in a sphere background.

This observation suggests that  the RT formula may be modified to correctly compute the holographic entanglement entropy of the \ttbar deformed theories.  We will address that in future work.

\section*{Acknowledgments}
I thank Huajia Wang for the valuable discussion and collaboration on related topics.  I also want to thank many of the members of KITS for interesting related discussions. 
JT is supported by  the National Youth Fund No.12105289 and funds from the UCAS program of special research associate.

\appendix
\section{Convention and Notation}
\label{Convention}
\renewcommand{\theequation}{A.\arabic{equation}}
\setcounter{equation}{0}
We are interested in 2d Euclidean spacetime and we define the holomorphic and anti-holomorphic coordinates as
\bea 
z=x+\im y,\quad \bar{z}=x-\im y.
\eea 
The partition function is defined via the path integral
\bea 
Z=\int \mathcal{D}[\phi(x,y)]e^{-S[\phi(x,y)]},
\eea 
where $\phi(x,y)$ denotes all the dynamical fields and $S[\phi]$ is the action. Assuming that the background metric is $ds^2=g_{\alpha\beta}dx^\alpha dx^\beta$ then the energy-momentum tensor is defined as
\bea 
T^{\alpha\beta}=-\frac{2}{\sqrt{g}}\frac{\delta S}{\delta g_{\alpha\beta}}.
\eea 
In the coordinate of $(z,\bar{z})$, the energy-momentum tensor is
\bea 
&&T_{zz}=\frac{1}{4}\(T_{11}-T_{22}-\im T_{12}-\im T_{21}\),\\
&&T_{\bar{z}\bar{z}}=\frac{1}{4}\(T_{11}-T_{22}+\im T_{12}+\im T_{21}\),\\
&&T_{z\bar{z}}=T_{\bar{z}z}=\frac{1}{4}\(T_{11}+T_{22}\)\equiv\Theta.
\eea 
For later convenience, we also introduce the renormalized energy-momentum tensor
\bea 
T=-2\pi T_{zz},\quad \bar{T}=-2\pi T_{\bar{z}\bar{z}}.
\eea 
The \ttbar composite operator is defined as
\bea 
\text{\ttbar}\equiv \frac{1}{8}\(T_{\alpha\beta}T^{\alpha\beta}-(T^\alpha_\alpha)^2\)=T_{zz}\bar{T}_{\bar{z}\bar{z}}-\Theta^2.
\eea 
We will also assume \ttbar satisfies the factorization property  
\bea \label{factor}
\langle \text{\ttbar}\rangle=\frac{1}{8}\(\langle T^{\alpha\beta}\rangle \langle T_{\alpha\beta}\rangle-\langle T^\alpha_\alpha\rangle^2\),
\eea 
which is true for field theories living on infinite Euclidean planes and cylinders or in holographic CFTs. Then the deformed theory can also be defined through the flow equation \cite{McGough:2016lol,Shyam:2017znq}:
\bea \label{trace}
\langle T^\alpha_\alpha\rangle=\frac{c}{24\pi}R+\frac{\mu}{4}\(\langle T^{\alpha\beta}\rangle \langle T_{\alpha\beta}\rangle-\langle T^\alpha_\alpha\rangle^2\)
\eea 
together with the conservation equation $\nabla_\alpha\langle T^{\alpha\beta}\rangle=0$. 
\section{Sphere partition functions}
\renewcommand{\theequation}{C.\arabic{equation}}
\setcounter{equation}{0}
In this appendix we re-derive the sphere partition function in our convention following \cite{Donnelly:2018bef}. 
We consider the metric $ds^2=e^{\sigma}(d\theta^2+\sin^2\theta d\phi^2),\quad \sigma=2\log r$. Considering a small Weyl transformation
\bea 
&&\sigma \rightarrow \sigma+\delta \sigma,\quad e^\sigma\rightarrow (1+\delta\sigma)e^\sigma,\\
&& g_{\alpha\beta}\rightarrow (1+\delta\sigma) g_{\alpha\beta}
\eea 
the action changes as
\bea 
\delta S=-\frac{1}{2}\int \sqrt{g}\delta \sigma T^\alpha_\alpha d^2x \quad \rightarrow \quad \frac{\delta S}{\delta \sigma}=\frac{\delta S}{\delta r}\frac{\delta r}{\delta \sigma}=-\frac{1}{2}\int \sqrt{g}T^\alpha_\alpha d^2x
\eea 
therefore 
\bea 
\frac{d}{dr}\log Z=\frac{1}{r}\int d^2 x\sqrt{g}T_\alpha^\alpha.
\eea 
More generally if we include the boundary then 
\bea 
\delta \log Z=\frac{1}{2}\int_M d^2 x\sqrt{g}T_\alpha^\alpha\delta \sigma+\frac{c}{24\pi}\int_{\partial M}K \delta \sigma dl
\eea 
where $K$ is the geodesic curvature of the boundary. For the sphere, there are no boundaries so we do not need to include the boundary term. Next, we want to determine $T_\alpha^\alpha$ from the trace relation \eqref{trace}. The crucial observation is that by symmetry the stress tensor takes the form $T_{\alpha\beta}=\alpha g_{\alpha\beta}$. Substituting into the trace relation gives
\bea 
\alpha=\frac{\sqrt{\frac{c \mu }{6 \pi  r^2}+4}-2}{\mu }.
\eea 
thus 
\bea 
\frac{d\log Z}{dr}=\frac{8 \pi  r \left(\sqrt{\frac{c \mu }{6 \pi  r^2}+4}-2\right)}{\mu }=-\frac{dS_{\text{QFT}}^\mu}{dr}.
\eea 
Integrating both sides and imposing the initial condition $S_{\text{QFT}}^\mu(r=0)=0$ gives
\bea 
S_{\text{QFT}}^\mu&=&-\frac{i \pi  c \mu +4 \sqrt{6 \pi } r^2 \sqrt{\frac{c \mu }{r^2}+24 \pi }+2 c \mu  \tanh ^{-1}\left(\sqrt{\frac{c \mu }{24 \pi  r^2}+1}\right)-48 \pi  r^2}{6 \mu } \label{fieldsphere}\\
&=&-4\sqrt{\frac{2\pi c }{3\mu}}r+\frac{8\pi r^2}{\mu}+\mathcal{O}(r^3)
\eea 
or
\bea 
S_{\text{QFT}}^\mu=-\frac{c}{6}+\frac{c}{6}\log\frac{c\mu}{96\pi r^2}-\frac{c^2\mu}{576\pi r^2}+\mathcal{O}(\mu^2)
\eea 
which does not have a $\mu=0$ limit.

\end{document}